# Deep Learning-Based Automatic Detection of Poorly Positioned Mammograms to Minimize Patient Return Visits for Repeat Imaging: A Real-World Application

Vikash Gupta[1,†], Clayton Taylor[2,†], Sarah Bonnet[3], Luciano M. Prevedello[2], Jeffrey Hawley[2], Richard D White[1], Mona G Flores[4], and Barbaros Selnur Erdal[1]

[1]Department of Radiology, Mayo Clinic, Jacksonville, Florida
[2]Department of Radiology, Ohio State University, Columbus, Ohio
[3]Department of Radiology, West Virginia University School of Medicine, Morgantown, West Virginia
[4]Medical AI, NVIDIA Inc., Santa Clara, California
[†]Co-first authors.

**ABSTRACT**

Screening mammograms are a routine imaging exam performed to detect breast cancer in its early stages to reduce morbidity and mortality attributed to this disease. In order to maximize the efficacy of breast cancer screening programs, proper mammographic positioning is paramount. Proper positioning ensures adequate visualization of breast tissue and is necessary for effective breast cancer detection. Therefore, breast-imaging radiologists must assess each mammogram for the adequacy of positioning before providing a final interpretation of the examination; this often necessitates return patient visits for additional imaging. In this paper, we propose a deep learning-algorithm method that mimics and automates this decision-making process to identify poorly positioned mammograms. Our objective for this algorithm is to assist mammography technologists in recognizing inadequately positioned mammograms real-time, improve the quality of mammographic positioning and performance, and ultimately reducing repeat visits for patients with initially inadequate imaging. The proposed model showed a true positive rate for detecting correct positioning of 91.35% in the mediolateral oblique view and 95.11% in the craniocaudal view. In addition to these results, we also present an automatically generated report which can aid the mammography technologist in taking corrective measures during the patient visit.

**Keywords** Mammogram, Mammography positioning, Breast cancer, Deep learning, MQSA

## 1 Introduction

Breast cancer is the most common malignancy in women worldwide, alone accounting for an estimated 25% of cases[1]. Mammography-based screening programs have been shown to result in at least 30% decrease in breast cancer mortality[2]. Consequently, screening programs have been widely adopted; in the United States, approximately 40 million mammograms are performed annually[3]. Nevertheless, the efficacy of screening mammography relies on the use of proper technique. Poor breast positioning in mammography can result in undetected malignancy, as well as repeat examinations with added costs[4–9]. Inadequate positioning in mammography has been found to be the most common technical cause of undetected breast cancers[4,10].

In an effort to ensure that quality standards are met for mammography performed in the United States, the Food and Drug Administration began regulating mammography through the federal Mammography Quality Standards Act (MQSA) legislation. This legislation mandates that accrediting bodies must assess specific imaging attributes of the clinical images, including positioning. Inadequate breast positioning has been reported as the most common image-quality deficiency in mammography and the leading cause of mammography facility accreditation failures[11]. In facilities accredited by the American College of Radiology, 79% of accreditation failures in 2015 were secondary to errors in breast positioning[11]. The Enhancing Quality Using the Inspection Program (EQUIP) initiative, introduced by the Division of Mammography Quality Standards in 2017, expanded the MQSA inspection protocol to emphasize the importance of image quality and facility clinical image review.

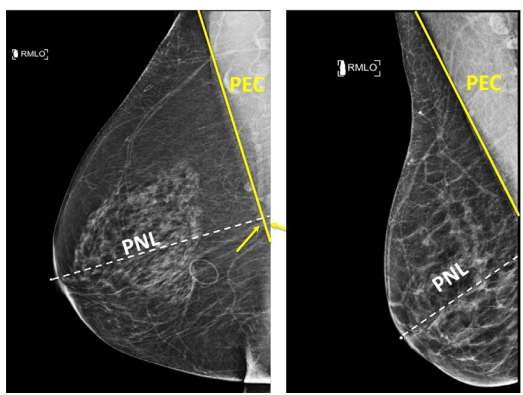

**Figure 1.** Example of adequately and inadequately positioned mediolateral-oblique (MLO) views. **Left:** In a well-positioned MLO view, the pectoralis muscle (PEC) line and posterior nipple line (PNL) intersect [arrow] within the image. This indicates that the image contains adequate coverage of posterior breast tissue. **Right:** In an inadequately positioned MLO view, the PEC line and PNL fail to intersect within the image domain.

The typical screening mammography workflow is standardized and involves a single imaging encounter with the acquisition of standard mammographic views in two positions, including the mediolateral-oblique (MLO) and cranial-caudal (CC) views of each breast. With the patient already having departed the imaging facility, the images are typically reviewed by the interpreting radiologist later that day in batches with results subsequently communicated to the patient by letter as required by MQSA regulations. In the process, the adequacy of breast positioning is assessed by the radiologist during the course of reviewing each mammographic image. If the study does not meet the necessary quality standards, the radiologist will request that the necessary mammographic images are repeated, which requires the patient to return to the facility for additional imaging. This process is inefficient for both the patient and the facility, precludes the interpreting radiologist from providing immediate constructive feedback to the mammography technologist, delays examination completion, and hinders the informing of patients and caring physicians about the screening results. The associated costs are manifested in financial expenses (e.g., patient return travel, repeat imaging related), lost time (e.g., patient away from work or family), and emotional toll (e.g., patient anxiety during result delay).

As previously discussed, adequate breast positioning is critical in mammography as it ensures that the maximum amount of breast tissue is visible on the clinical images, particularly the posterior breast tissue [Figure 1]. The adequacy of tissue included on the MLO view is assessed by ascertaining that the pectoralis muscle (PEC) line extends to or below a line drawn from the nipple perpendicular to the course of the PEC line, known as the posterior nipple line (PNL). If the PEC line does not extend to the PNL, there is a possibility of breast tissue exclusion from the mammogram.

The CC view is assessed by looking for visible fibers of the PEC (visible only in approximately 15% of mammograms), or more often by extending a line from the nipple perpendicular to the chest wall, also referred to as the posterior nipple line (PNL) [Figure 2]. If the length of the PNL on the CC view is within 1 cm of the PNL on the MLO view, then the CC view is deemed to include enough posterior tissue.

Given the importance of proper positioning for mammographic cancer detection, the large number of mammograms performed annually, and the inefficient workflow for identifying and remedying exams with poor positioning, the value of improving the process of identifying poorly positioned mammograms is clear. However, while much attention has been placed on deep learning and artificial intelligence based autonomous cancer-detection methods, there has been minimal emphasis on novel strategies to improve quality and positioning.

Deep learning and artificial intelligence have produced great results in the field of medical imaging and can aid radiologists and clinicians in decision-making processes. Unlike large datasets available in the computer vision community, such as ImageNet[12], Pascal-VOC[13], or Microsoft-COCO[14], medical imaging datasets are very small. In the case of such a small dataset, it is often useful to apply techniques, such as transfer learning[15]. In transfer learning, we choose a pre-trained neural network architecture and re-purpose it for the next similar task at hand. We can initialize a network under development with weights from a previously trained model on a larger dataset. Transfer learning has been successfully used to detect bone fracture on radiographs[16–18] and the presence vs. absence of coronary artery atherosclerosis in cardiac computed tomography angiography images[19]. These use-cases represent transfer learning-enhanced classification-based tasks. However, in the context of the present application, we use transfer learning for a regression task (i.e., to predict the PEC line and PNL).

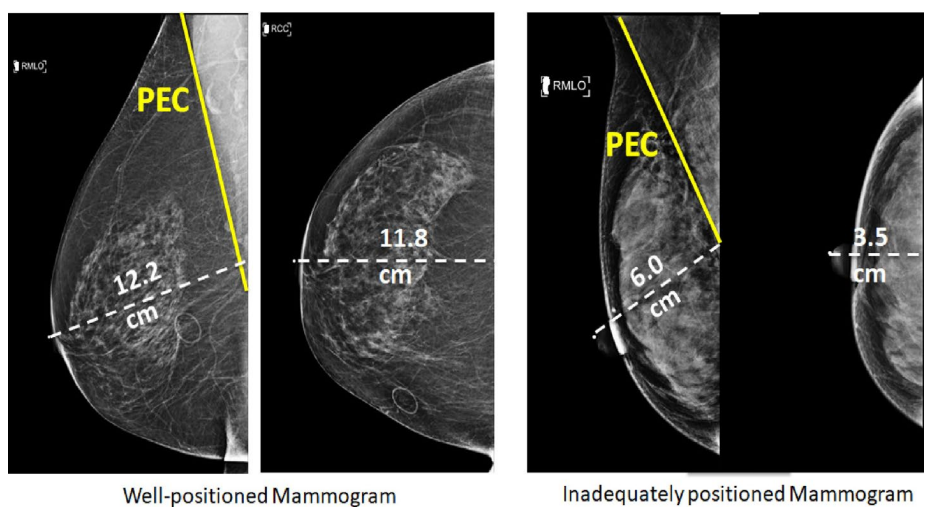

**Figure 2.** Relationship between the posterior nipple line (PNL) on mediolateral-oblique (MLO) and cranial-caudal (CC) views for well-positioned and inadequately positioned mammograms. **Left:** The difference between the PNL lengths in the MLO and CC views should be less than 1 cm, as in this case. **Right:** If the difference between the PNL lengths in the MLO and CC views is more than 1 cm, this indicates that the CC view may be inadequately positioned, leading to the exclusion of posterior breast tissue.

In the present report, we present a step-by-step process for characterizing the positioning of the breast. This includes descriptions of each algorithmic piece and how the mammography technologist can determine the exact reason for any inadequate positioning immediately upon completion of imaging with the still patient present for any needed repeat imaging.

## 2 Methods

While our ultimate goal was to evaluate the positioning on a mammogram, we first needed to develop our deep learning algorithm for PEC line and PNL predictions. We also needed to develop a separate algorithm for detecting the radiopaque marker (i.e., BB), which is essential for predicting the PNL on the CC view. Only after achieving these key tasks, we could use the prediction results to evaluate the positioning of mammograms. The data separation for each of these tasks is shown [Figure 3], followed by a discussion of each component in the appropriate context.

All screening mammogram studies were performed using 2D full-field digital mammography (FFDM) [Hologic: Marlborough, MA] in a dedicated breast center, at satellite imaging facilities, or in mobile mammography units. The standard protocol for our practice during this period included obtaining 2D FFDM images for each screening mammographic view even when 3D tomosynthesis images were obtained. For tomosynthesis cases, only the 2D FFDM images were utilized in this study.

### 2.1 Data sets

This retrospective study was approved by the local Institutional Review Board (Ohio State University), with a waiver of the requirement to obtain informed patient consent. Screening mammograms performed between January 2012 and July 2017 were mined in the PACS system using keywords indicating issues with positioning; ultimately, each of the 205 selected mammogram studies were determined to contain at least one image considered by a dedicated breast imaging radiologist to be technically inadequate due to mal-positioning. Subsequent case-exclusion criteria included: 1. age under 30 years; 2. prior breast conservation/reduction surgery; 3. breast piercings; or 4. breast implants; they led to the exclusion of only 11 patients.

The remaining 194 screening mammogram studies yielded a total of 508 MLO, and 379 CC repeat mammographic images during repeat visits [Figure 3]; in each case, multiple images were often acquired in order to include all breast tissue or to address another mal-positioning issue. To obtain additional normal cases with appropriate breast positioning, a consecutive series of 133 screening mammogram studies from the same time period and deemed to be technically adequate at initial interpretation and containing exactly four images (a single MLO and CC for each breast) were also included.

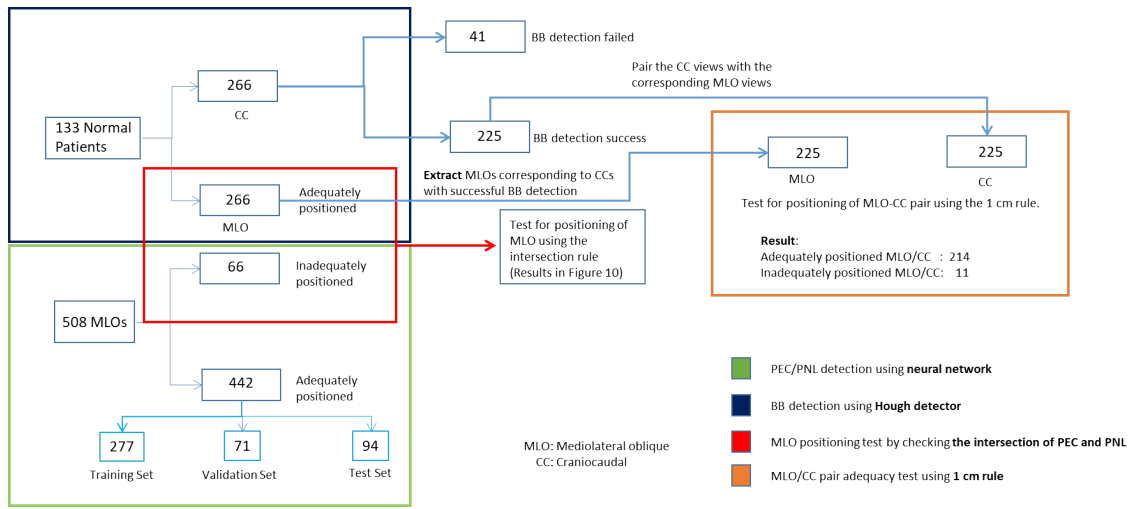

**Figure 3.** Algorithm-development populations and datasets. The 133 normal patients, providing 266 well-positioned MLO views and CC views, are shown (top left: dark blue box). Of the 508 MLO views from the 194 patients with mal-positioned views included in their studies, 442 adequately positioned images were used for PEC/PNL line-prediction (bottom left: light green box). Data separation for further evaluations: MLO adequate positioning test using the PEC/PNL line-intersection rule (red box), BB-detection test using Hough transform, positioning of MLO and CC images using the 1-cm rule (orange box)

Of the 508 labeled MLO images, the research review process identified 442 images that were considered technically adequate [Figure 3]. These were used for training the neural network for predicting the PEC line and PNL; for that purpose, they were divided into training (n=277), validation (n=71), and test (n=94) sets. The remaining 66 MLO images initially identified as poorly positioned only because of a lack of posterior tissue representation (manifested by PEC line and PNL not seen to intersect) were kept for testing the positioning algorithm performance according to MQSA standards. These 66 poorly positioned MLO views were tested against the 266 correctly positioned MLO views from the 133 normal cases.

## 2.2 Data labeling

Annotating radiological images is a challenging task[20]. Unfortunately, commonly used clinical imaging viewers are often unsuitable for labeling images for training AI algorithms. As shown [Figure 4], the PEC line and PNL, as well as the radiopaque tags, are marked on both MLO and CC views using LabelMe[21], a publicly available tool for labeling images. A radiologist involved in this study drew straight lines demarcating the PEC line on MLO views. The PNLs were also drawn on both MLO and CC views. In the case of MLO views, the PNL was aligned approximately perpendicular to the PEC line. However, in the case of CC views, the PNL was approximately a horizontal line starting at the BB marker and reaching the end of the image either on the left or right, depending on whether the left or right breast was under consideration. Last, the radiopaque label indicating the image view was annotated using a box region-of-interest (ROI). The ROI is defined by two coordinate points on the diagonal corners. LabelMe stores the annotations in a Javascript Object Notation (JSON) format in separate files for each image.

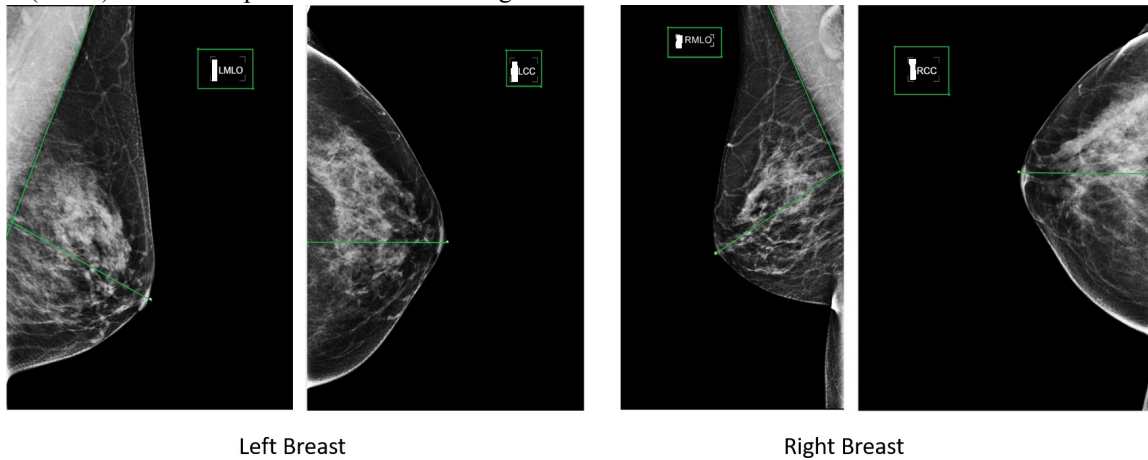

Left Breast  Right Breast

**Figure 4.** An example of tagging performed under joint consensus review by radiologists to establish ground truth. Labels

for both views (MLO and CC) of the left and right breast are indicated. Both the PEC line and PNL are shown on MLO views; on the CC view, only the PNL is shown. An ROI delineates the image view used for each image.

## 2.3 Data augmentation

Deep learning algorithms typically employ standard data augmentation techniques to increase the size and variability in the dataset. The traditional augmentation techniques used in deep learning involve image flipping, random rotations, translations, cropping, shearing, and elastic deformations. Of these, we applied transformations, including random horizontal flips, to introduce lateral (left or right) invariance. Random rotations about the center of the image were also used, with overall resizing of the rotated images in order to overcome cutting-off of an object of interest (in this case, the PEC line) due to the expansion of the breast representation if overall image-size is fixed [Figure 5]. The same transformations were applied to the image annotations. We did not use translation or cropping as augmentation techniques, as they remove part of the object of interest (i.e., PEC line and PNL) from the image.

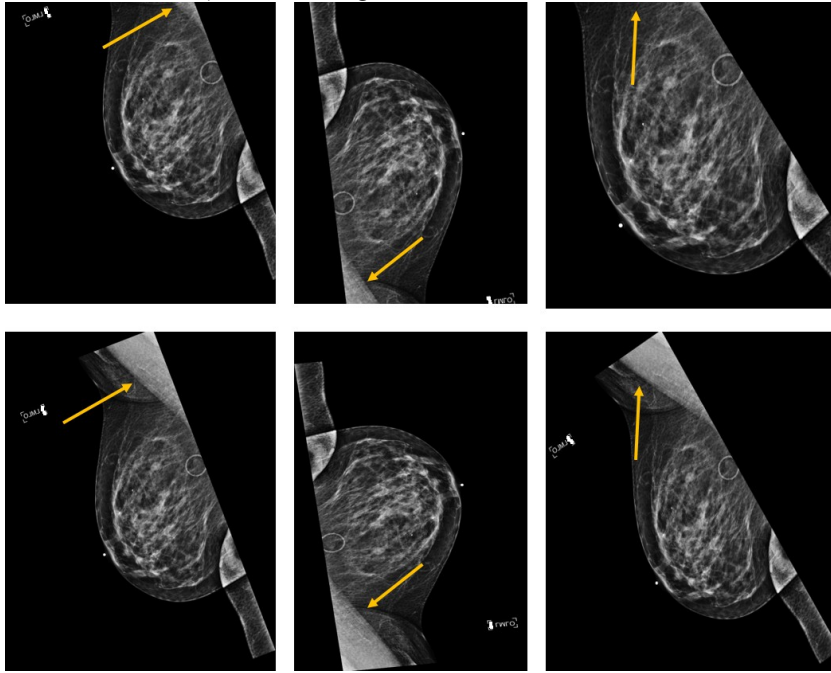

**Figure 5.** Rotated views need correction for expansion of breast representation size. Image rotation without compensation for expanded breast representation can cause cropping of important imaging landmarks [arrows] (top row). This is corrected by increasing the overall image size (bottom row).

## 2.4 Predicting the PEC line and PNL on MLO view

A line is defined by the Cartesian Coordinates (x, y) of its two end-points. For predicting the PEC line and PNL, we trained a neural network to predict these two points. As the tasks and the input/output are very similar, neural networks with the same architecture were used for both. In the following sections, we discuss the neural network architecture used for predicting the PEC line and PNL, as well as the parameters and loss-functions used for training.

**Neural Network:** Training deep neural networks is often challenging when the dataset size is fairly limited. In general, whenever the dataset sizes are small, transfer learning is the preferred approach. Transfer learning is a deep learning technique that leverages prior knowledge or information learned from larger datasets and fine-tunes the network for the dataset at hand. While the earlier layers of the network learn the higher level (coarse) features of an image, as shown by Zeiler et al[22], deeper layers learn the finer features of the image. In this project, we chose Inception-V3[23] as the base network; it is a neural network designed for image classification whose last layer contains 1000 fully connected nodes for each of the classes. We replaced the last layer with a single output node, which produces a vector with four elements corresponding to the two ordinates (*x* and *y*) of the two output points [Figure 6]. We initialized the network with the pre-trained weights from the ImageNet dataset. Instead of using the cross-entropy as the cost function, we used the logarithm of hyperbolic cosine (Log-Cosh) as the cost function to be optimized. The cost function is defined as:

$$L(y^{true}, y^{pred}) = \sum_{i=1}^{n} \log[\cosh(y_i^{true} - y_i^{pred})]$$

where $L$ is the loss value, $y^{true}$ and $y^{pred}$ are the true and the predicted values, respectively. The Log-Cosh loss takes on the behavior of squared loss when the cost function is small and acts as an absolute loss when the cost function is large. This behavior is similar to Huber loss, but the cost function has the advantage of being twice differentiable everywhere.

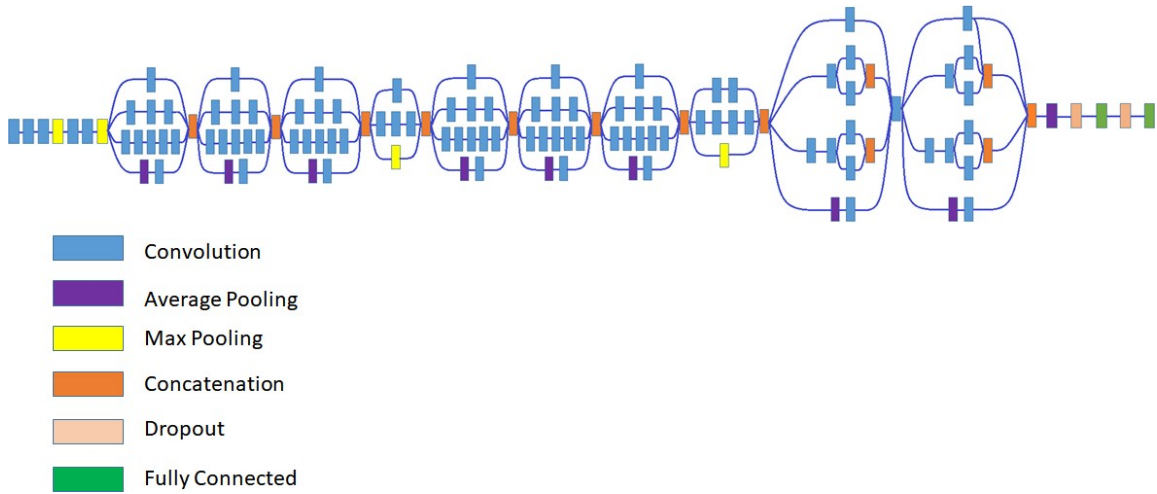

**Figure 6.** Transfer learning using the Inception-V3[23] neural network. The final layers are modified with a fully connected layer with linear activation. The last fully-connected layer [on the right: green] contains a single node that produces a vector of four outputs as opposed to a single softmax output, as in the case of Inception-V3 network.

The average size of a mammogram is approximately 3500 pixels in width and 4000 pixels in height. The images and annotations were resampled to 250 × 250 pixels. In addition, the image intensities were normalized to a range (0, 1). We used a batch size of 12 images and a learning rate of 0.0001 for training, which continued for 150 epochs. During the training process, we monitored the validation loss and saved the model, which achieves the lowest validation loss. Each epoch took approximately 378 seconds and the total training process required approximately 16 hours to complete using a single graphics processing unit (one V100 on DGX1 system: NVIDIA, Inc., Santa Clara, CA).

## 2.5 Detecting the BB (nipple) and PNL on CC view

A radiopaque marker (i.e., BB) is placed over the nipple to assign landmarks on the mammogram. The BB marker is a hyper-intense circular blob, approximately 20 pixels in diameter. The radiopaque marker is known to have the highest intensity in the image. As we know that the shape of the BB marker is circular, we can use the Hough Circle detector[24]. Specifically, we used the OpenCV implementation, *HoughCircles*[25]. The circle detection algorithm takes inputs as the minimum and maximum radii of the circles we intend to detect. We found 10 and 20 as the most optimal minimum and maximum radii for the image. The output of the Hough-Circle transform is a set of center-points and the radii of the detected circles. The circle detector will yield multiple potential BB candidates [Figure 7]. We can use the fact that the BB has a uniform intensity for filtering out the false-positive candidates for BB detection. This filtering process is achieved using the 9-way connectivity rule. We checked if all 9 pixels connected to the center point of the circle have the same intensity values. Only the candidate passing this connectivity rule is the BB marker. Once the BB is detected, we drew a horizontal line to either the left or right extremity of the image, depending on the laterality of the breast. The length of the line segment gives the PNL length on the CC view ($d_{CC}$) of the breast.

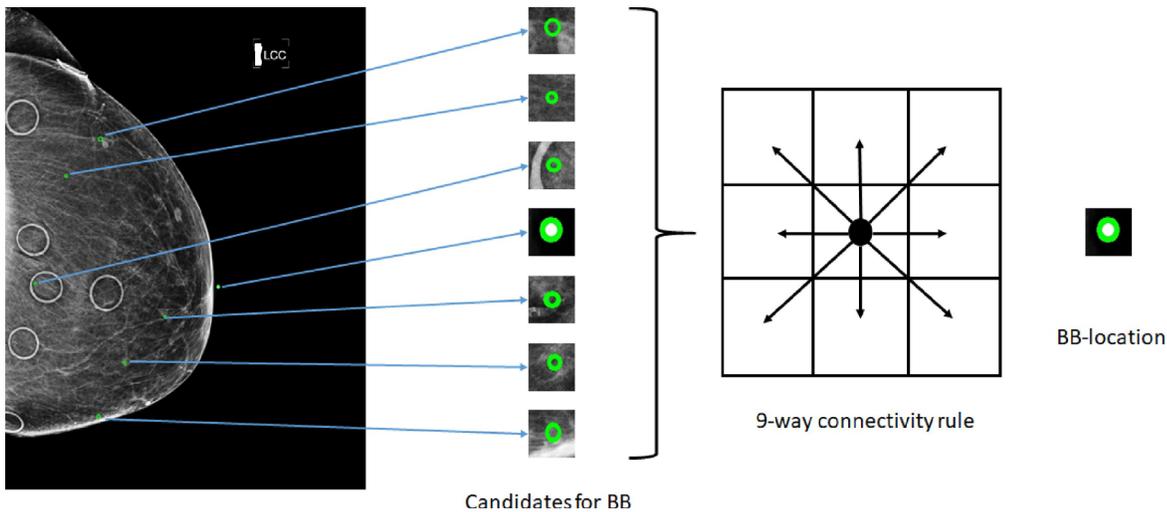

**Figure 7. BB detection:** A pipeline showing the algorithm for BB detection. The image on the left shows all possible candidates for the BB location. Each of the candidates is then subjected to a 9-way connectivity rule; all the pixels in the vicinity of the center point should have the same intensity as that of the center. Only the BB satisfies this standard.

### 2.6 Algorithm: Predicting the positioning of MLO and CC views

Once the PEC line and PNL are predicted on the MLO and CC views, the positioning of individual views can be determined. For the MLO view to be considered adequate, the PEC line and PNL must intersect within the image bounds; otherwise, the MLO view is inadequate and likely under-represents posterior tissue. Only if the point of intersection lies within the image boundaries can we calculate the PNL length ($d_{MLO}$) and make predictions about the adequacy of the CC view. The PNL length on the CC view ($d_{CC}$) is then calculated. The absolute difference between $d_{MLO}$ and $d_{CC}$ should be less than 1 cm for the CC view to be considered to be adequate. These steps are outlined in Algorithm 1.

---

**Algorithm 1** Algorithm to compute decision making on mammogram positioning

---

**Input:** MLO view, CC view of the mammogram
**Output:** Decision on mammogram positioning
   **Step 1:** Predict the PEC line and PNL on MLO view.
   **Step 2:** Check MLO positioning (intersection rule).
      a. **If** PEC line and PNL intersect within the image boundaries, then MLO is adequately position and **accept the view.**
      b. **Else,** MLO is inadequately positioned and **reject the view.**
   **Step 3:** Predict BB location on the CC view.
        Draw the PNLs on the CC view.
   **Step 4:** Calculate length of PNLs on CC ($d_{cc}$) and MLO ($d_{MLO}$) views.
        Calculate $d_{diff} = |d_{cc} - d_{MLO}|$
      a. **If** $d_{diff} < 1$ cm, **accept the view**.
      b. **Else,** the CC is inadequately positioned and **reject the view.**
**Exam complete**

---

## 3. RESULTS

### 3.1 Predicting PEC line and PNL on MLO

In order to predict the PEC line and PNL on the MLO view, the following occurs:
1. the image is down-sampled to a size of 250 X 250 pixels, and the intensity is normalized to 0 and 1. The trained model is used to make the prediction on the down-sampled images. A comparison between the ground-truth labels and the predicted labels on these images is shown [Figure 8].
2. the images of the predictions are up-sampled to the original image resolution using linear interpolation. Plots show the distribution of errors in the end-points for the PEC line and PNL across the test set [Figure 9]. In addition, we show a comparison between the lengths of the predicted PNL and the ones drawn by the radiologists [Figure 10].

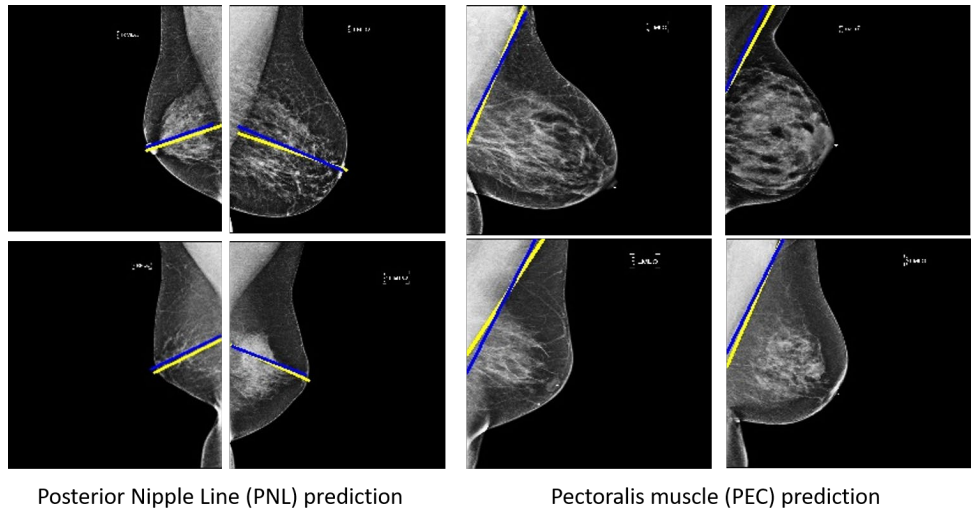

Posterior Nipple Line (PNL) prediction                    Pectoralis muscle (PEC) prediction

**Figure 8.** Predicting the pectoralis muscle (PEC) line and the posterior nipple line (PNL) on the down-sampled images. [predicted line: yellow; ground-truth line: blue]

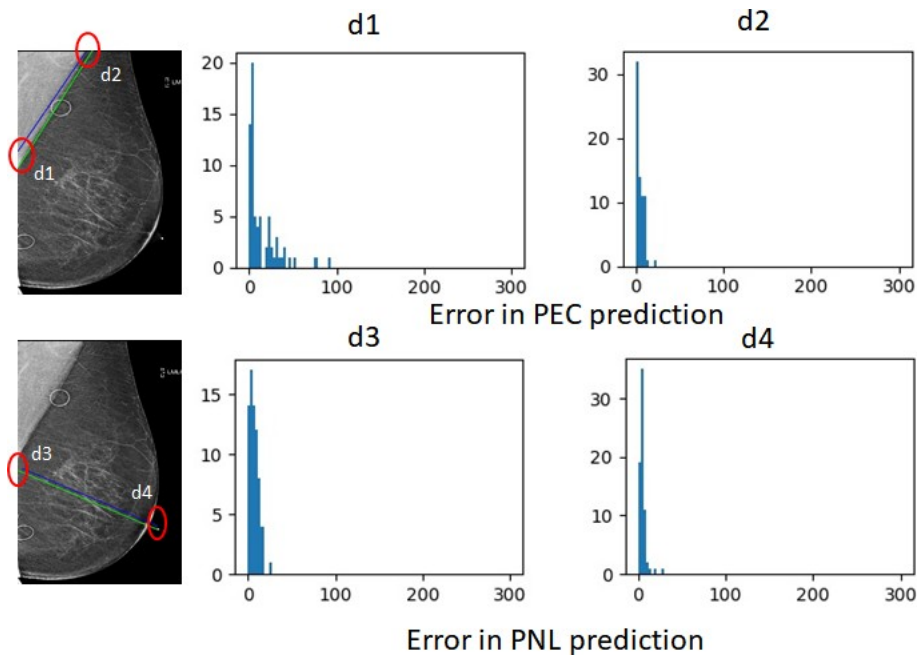

**Figure 9.** Errors on the end-points while predicting the pectoralis muscle (PEC) line and the posterior nipple line (PNL). The errors are measured based on the end-points of the PEC line and the PNL line. In the adjoining histograms on the right, the error distribution for all the four points in the test set is shown. [predicted line: blue; ground-truth line: green]

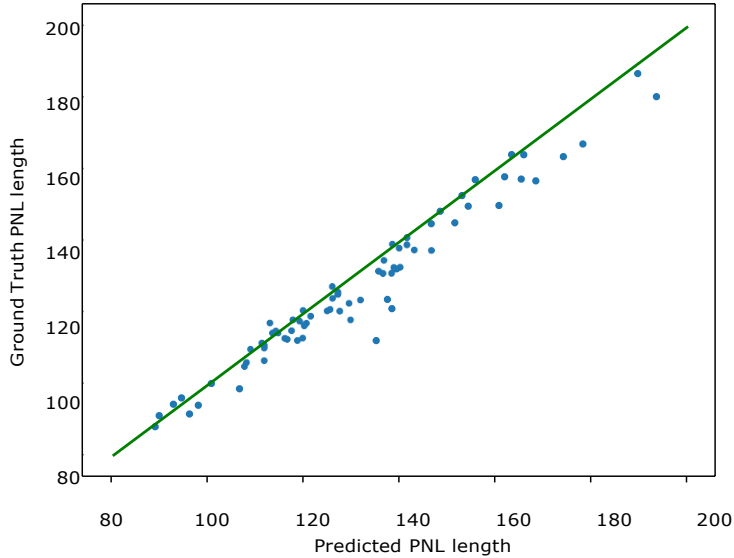

**Figure 10.** Comparison between the predicted lengths of the posterior nipple line (PNL) and the ground truth.

## 3.2 Predicting MLO positioning

After predicting the PEC lines and PNL on an MLO view, we can calculate their point of intersection and the length of the PNL on the MLO view ($d_{MLO}$). As per the MQSA standards, this point of intersection should lie within the boundaries of the image, and the PNL length is calculable, for the MLO view to be acceptable. From the confusion matrix for the proposed intersection rule [Figure 11], the proposed method is very accurate in detecting adequately positioned MLO views, based on a true-positive rate of 91.35%. On the other hand, the method is not particularly useful in detecting a poorly positioned MLO, with true-negative rate of only 45%.

|  | Predicted positioning | | |
|---|---|---|---|
| N = 332 | Adequate | Inadequate | Total |
| Adequate | 243 | 23 | 266 |
| Inadequate | 36 | 30 | 66 |

(Ground Truth on left axis)

**Figure 11.** Confusion matrix showing the efficacy of prediction of MLO positioning based on the intersection of PEC line and PNL.

## 3.3 Predicting CC positioning

The positioning on the CC view is judged based on its corresponding MLO view. For the CC view, we first need to predict the BB. If a BB is not correctly detected, we cannot make any prediction about the PNL length on the CC view, and thus about its positioning adequacy. Our proposed algorithm was able to detect the correct BB location in 84.5% (225/266) of cases [Figure 3].

We can only predict the adequacy of the positioning for the CC view if we can detect the BB accurately. Thus, in order to test the efficacy of the MQSA 1-cm rule, we only included the CC images, where the correct BB was predicted. Out of 225 adequately positioned MLO/CC pairs, we were able to accurately predict the adequate positioning for 214 MLO/CC pairs and failed to do so for the remaining 11 MLO/CC pairs [Figure 3], leading to a true positive rate of 95.11%.

## 3.2 Generating an automated report on breast positioning: Real-world application

In clinical practice, we are able to generate a text report, along with images with the predicted PEC line and PNL drawn.

The goal of this output is to provide enough information to the mammography technologist about breast positioning. If the technologist sees that the predicted PEC line and PNL are not correct, or the BB is not detected, he/she can disregard the text report. Here we will discuss two examples of automatically generated reports.

In the first example [Figure 12], we show a report generated for a patient with four scans. The points of intersection of the PEC line and PNL on both left and right MLO views (LMLO and RMLO, respectively) are shown, along with corresponding PNL length. Similarly, PNL lengths are calculated on the CC views, and the difference between the two lengths is reported. Both MLO and CC views are adequate for the left breast. However, for the right breast, while the PEC line and PNL intersect on the MLO view (i.e., adequate), the difference in PNL lengths between the two views for the right breast is more than 1 cm. Thus, the right CC view (RCC) is inadequate. This actionable, targeted result is available to the mammography technologist while the patient is still in the imaging facility. The technologist can themselves immediately make a judgment if the patient needs to undergo reimaging for "only" the right CC view.

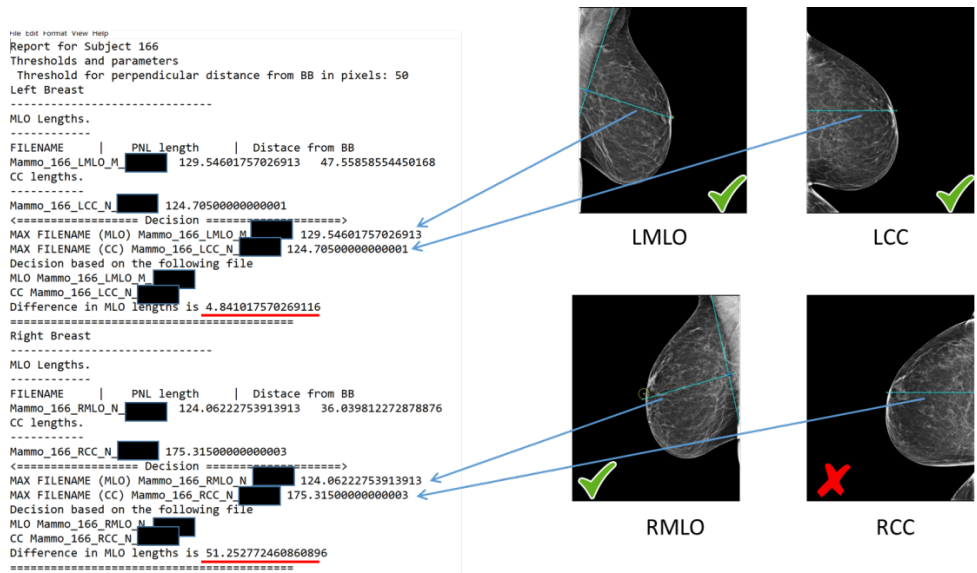

**Figure 12.** Automated report on breast positioning. The PEC line and PNL intersect on both left (LMLO) and right (RMLO) views. The difference between the lengths of the posterior nipple line is reported.

In the second example (Figure 13), we show the report generation for a patient where the proposed algorithm failed to make any decisions. For the left breast, the mammography technologist can see that the PEC line and PNL are not intersecting on the MLO (LMLO) view within the image boundaries. He/she can then use their judgment on the positioning of the image. Our algorithm failed to detect the BB on the CC view (LCC) of the image. Thus, no decision was made on the positioning of the left breast. However, in the case of the right breast, multiple MLO (RMLO) and CC (RCC) views are available. Out of the two RMLO views, the PEC line and PNL do not intersect in the left, but do intersect on the right image. The PNL length is calculated based on the image in which they intersect. For the two CC views (RCC), we consider the one with a longer PNL because it ensures more breast tissue to be present. Finally, the difference in the PNL length is computed, and a decision is reported for the right breast positioning.

## 4  Discussion and future work

In this report, we have presented a deep learning-based method for determining if the breast is well-positioned on a mammographic images. The proposed method is based on the MQSA standards for breast positioning. We showed that by combining techniques from deep learning and feature-based machine learning algorithms, it is possible to accurately predict if the breast is adequately positioned. In addition, we showed that this task could be broken down into separate problems: 1. Predicting the PEC line and PNL on the MLO view, and 2. Predicting the PNL line on the CC view. Once the PNLs on both the MLO and CC views are predicted, we calculate their lengths and the absolute difference between them, which is finally used for predicting if the breast was well-positioned.

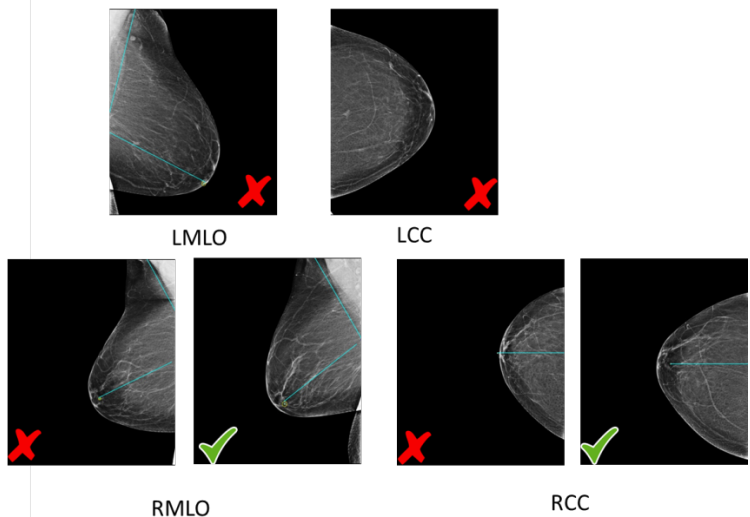

**Figure 13.** Automated report on breast positioning based on multiple images. The point of intersection on the LMLO view is outside the image boundaries, and no BB is detected on the LCC view. Thus, the algorithm did not make any decisions regarding the left breast. For the right breast, multiple MLO and CC views were present. The MLO, where PEC and PNL intersect within the image boundaries (RMLO-right) is selected. The CC view with the longest PNL (RCC-right) is selected.

Thus, we have presented a deep learning-based approach for predicting the PEC line and PNL on the MLO view. In addition, we also presented a method to predict the location of a BB nipple marker, which is then used to calculate the PNL length on the CC view. Combining the results from these separate problems, we can predict if the breast is adequately positioned. We showed that our proposed method has a high true-positive rate for detecting the adequately positioned mammograms. However, the true-negative rate for detecting the inadequately positioned mammograms is low.

However, we cannot necessarily conclude that this method is not a good alternative for detecting poorly positioned MLOs. There are several reasons to support this claim. First, the number of poorly positioned MLO images in our dataset is small compared to the number of normal images. Hence, it is difficult to make a statistically relevant claim. As the positioning algorithm is dependent on the guidelines proposed by MQSA, the errors are also dependent on the accuracy of PEC line and PNL detection. Further, as our neural network was trained on normal MLO views only, the PEC/PNL detection accuracy is relatively lower in the case of poorly positioned MLOs. We can improve on this aspect of the proposed work by collecting more mal-positioned images and including them in re-training our model.

One of the significant contributions of this work is the generation of an automated report about breast positioning. The report is designed to provide the mammography technologist with precise information about the quality of the positioning. Instead of delivering a binary result on breast positioning, it presents more actionable information to the technologist. For example, if the technologist sees that the predicted PEC line and the PNL in the MLO do not intersect, they can re-position the breast and repeat the MLO view. Similarly, if the PNL length on the CC mammogram is not within one centimeter of the PNL length in MLO, the technologist immediately has evidence that the CC image is inadequate. This type of actionable information will likely reduce both the time spent by the technologist determining image quality, and the number of poorly positioned exams necessitating repeat visits by patients to the imaging facility. In the future, we would like to implement this method to create a real-time feedback mechanism in a clinical environment

## 5 Acknowledgments

The authors would like to thank Dr. Ratnesh Kumar for his invaluable support and insights during the course of this project. We will also like to acknowledge that the data collection and algorithm development was performed in the Department of Radiology at The Ohio State University.